\documentclass{article}

\usepackage{mathrsfs}
\usepackage{amsfonts}
\usepackage{amsmath}
\usepackage{amssymb}
\usepackage{graphicx}
\newtheorem{Conjecture}{Conjecture}

\title{Extending cosmological natural selection}
\author{Gordon McCabe}

\begin{document}

\maketitle

\begin{abstract}

The purpose of this paper is to propose an extension to Lee Smolin's hypothesis
that our own universe belongs to a population of universes evolving by natural
selection. Smolin's hypothesis explains why the parameters of physics possess
the values we observe them to possess, but depends upon the contingent fact
that the universe is a quantum relativistic universe. It is proposed that the
prior existence of a quantum relativistic universe can itself be explained by
postulating that a process of \emph{cosmogenic drift} evolves universes towards
stable (`rigid') mathematical structures.

\end{abstract}

\section{Introduction}

According to current mathematical physics, there are many aspects of our
physical universe which are contingent rather than necessary. These include
such things as the values of the numerous free parameters in the standard model
of particle physics, and the parameters which specify the initial conditions in
general relativistic models of the universe. The values of these parameters
cannot be theoretically derived, and need to be determined by experiment and
observation. It transpires that the existence of life is very sensitively
dependent upon the values of these parameters (see Barrow and Tipler 1986, for
a compendious survey). If a universe had values for these parameters only
slightly different from the values they possess in our own universe, then that
universe would be incapable of supporting life. Hence, there is a need to
explain why a life-supporting universe exists.

In fact, the problem posed by the contingent values of the free parameters can
be generalised. If our physical universe is conceived to be an instance of a
mathematical structure, (i.e., a structured set), then it is natural to ask why
this mathematical structure physically exists and not some other. The
instantiation of one particular mathematical structure is contingent, and
requires an explanation.

One response to this problem of contingency is to postulate the existence of a
collection of universes, which realise numerous different mathematical
structures, and numerous different values for the parameters of physics. It is
common these days to refer to such a collection as a `multiverse'. Multiverses
can be distinguished by whether or not some physical process is suggested to
account for their existence. For example, Linde's chaotic inflation theory
(1983a and 1983b), and Smolin's theory of cosmological natural selection
(1997), both postulate the operation of physical processes which yield
collections of universes, (or causally disjoint `universe-domains', in the case
of Linde's theory). Other multiverse proposals postulate universes which are
either not the outcome of a common process, or not the outcome of any process
at all (Tegmark 1998, 2008).

As Tegmark points out, all such proposals which suggest that ``some subset of
all mathematical structures\dots is endowed with\dots physical existence,"
(1998, p1), fail to explain why some particular collection of mathematical
structures is endowed with physical existence rather than another. Tegmark's
own response was to suggest that \emph{all} mathematical structures have
physical existence.\footnote{More recently, however, Tegmark (2008) has
incorporated the implications of G\"odel incompleteness and Church-Turing
uncomputability, by considering the possibility that only computable
structures, or finite computable structures, physically exist (2008, p22).} The
weak anthropic principle similarly postulates the existence of a collection of
universes which is sufficiently large and varied that the conditions which
permit the existence of life will be realised in at least some of the
universes. Both types of proposal accept that a life-permitting universe is a
highly untypical member of the universe collection, and both types of proposal
are difficult, if not impossible, to empirically test. In contrast, Smolin's
proposal of cosmological natural selection explains the existence of our
life-supporting universe, renders such universes highly typical, and is subject
to empirical test. We now proceed to expound Smolin's hypothesis.

\section{Cosmological natural selection}

To understand Smolin's idea, it is first useful to appreciate that the
conditions for natural selection to occur can be precisely defined formally, in
complete abstraction from any particular physical instance. If a collection of
physical systems satisfies these conditions, then that collection will, with
overwhelming likelihood, evolve by natural selection, irrespective of what
those systems are. The Darwinian evolution of biological systems by natural
selection, is just one particular case of this.

John Barrow asserts that natural selection (or, as he calls it,
`Darwinian evolution'), ``has just three requirements:

\begin{itemize}
\item{The existence of variations among the members of
a population. These can be in structure, in function, or in
behaviour.}
\item{The likelihood of survival, or of reproduction,
depends upon those variations.}
\item{A means of inheriting
characteristics must exist, so that there is some correlation
between the nature of parents and their offspring. Those
variations that contribute to the likelihood of the parents'
survival will thus most probably be inherited.}
\end{itemize}

It should be stressed that under these conditions evolution is not
an option. If any population has these properties then it must
evolve," (Barrow 1995, p21).

Smolin hypothesises that there exists a population\footnote{Hereafter, a
collection of universes which are related in some way, will be referred to as a
`population' of universes.} of universes, and that the values of the free
parameters in the standard model of particle physics are variable
characteristics of the universes in the population. For simplicity, let us
accept that the values of the fundamental parameters of physics are fixed in
each universe, but can vary from one universe to another.

Smolin hypothesises that certain types of universe in the
population are reproductively active. He suggests that in those
universes where black holes form, a child universe is created
inside the event horizon of the black hole. Specifically, Smolin's
proposal is that ``quantum effects prevent the formation of
singularities, at which time starts or stops. If this is true,
then time does not end in the centers of black holes, but
continues into some new region of space-time\dots Going back
towards the alleged first moment of our universe, we find also
that our Big Bang could just be the result of such a bounce in a
black hole that formed in some other region of space and time.
Presumably, whether this postulate corresponds to reality depends
on the details of the quantum theory of gravity. Unfortunately,
that theory is not yet complete enough to help us decide the
issue," (1997, p93).

In the decade since Smolin proposed his idea, loop quantum gravity has made
some significant progress, and its application to cosmology now appears to
support Smolin's hypothesis. For example, in a recent review, Ashtekar asserts
that ``In the distant past, the [quantum] state is peaked on a classical,
contracting pre-big-bang branch which closely follows the evolution dictated by
Friedmann equations. But when the matter density reaches the Planck regime,
quantum geometry effects become significant. Interestingly, they make gravity
\emph{repulsive}, not only halting the collapse but turning it around; the
quantum state is again peaked on the classical solution now representing the
post-big-bang, expanding universe," (2006, p12). Nevertheless, it is fair to
say that the occurrence of such a bounce inside a black hole remains highly
speculative.

Smolin postulates that the reproduction which takes place is
reproduction with inheritance. He assumes that ``the basic forms
of the laws don't change during the bounce, so that the standard
model of particle physics describes the world both before and
after the bounce. However, I will assume that the parameters of
the standard model do change during the bounce," (1997, p94).
Smolin postulates that a child universe inherits almost the same
values for the parameters of physics as those possessed by its
parent. He postulates that the reproduction is not perfect, that
small random changes take place in the values of the parameters.
Hence, Smolin postulates reproduction with inheritance and
mutation. As Shimony puts it, ``the variable entities are
universes, and the theatre in which the variation occurs is
governed by the principles of quantum gravity (as yet not fully
constructed) and the form of the standard model," (1999, p217).
Smolin's scenario cannot explain why our universe is relativistic
rather than non-relativistic, and it cannot explain why our
universe is a quantum universe rather than a classical universe,
because the occurrence of black holes requires a relativistic
universe, and the occurrence of a `bounce' inside the horizon of a
black hole requires a quantum universe.

The number of black holes in a universe is determined by the
parameters of physics, hence the values of the parameters in a
universe determine the number of children born to that universe.
If Smolin's postulate that child universes are created inside
black holes with small random parameter mutations is indeed
correct, then a population that contains some black hole producing
universes, will probably evolve by natural selection. In
particular, a population with an exhaustive, initially uniform
distribution of parameter value combinations, will come to be
dominated by universes that maximise the production of black
holes.

In addition to the hypothesis that there is a population of
universes evolving by natural selection, Smolin suggests that the
parameter values which maximise black hole production, and
therefore child universe birthrate, are also the values which
permit the existence of life. If the universe types with the
highest birthrate are also those universes which permit life, then
universes which permit life will come to dominate the population
of universes.

The hypothesis that there is a population of universes evolving by
natural selection is distinct from the hypothesis that the
parameter values which maximise black hole production are the same
parameter values which permit life. One hypothesis could be true,
and the other false. Only if both are true will life-permitting
universes come to dominate the population of universes. If child
universes were created inside black holes with small random
parameter variations, but the parameter values which maximise
black hole production were \emph{not} the same parameter values
which permit life, then there would be a population of universes
which evolves by natural selection, but in which life-permitting
universes do not come to dominate the population.

A weak anthropic principle explanation that imagines a collection of unrelated
universes, rather than a population of universes evolving by natural selection,
holds that life-permitting universes are \emph{special} members of the
collection. In contrast, Smolin's dual proposal that (i) there is a population
of universes evolving by natural selection, and (ii) the parameter values which
maximise black hole production are the same parameter values which permit life,
holds that life-permitting universes are \emph{typical} members of the
collection. In terms of carbon and organic elements, for example, the theory of
cosmological natural selection ``predicts that our universe has these
ingredients for life, not because life is special, but because they are typical
of universes found in the collection," (Smolin 1997, p204).

Vilenkin (2006) has recently framed an argument which poses a serious challenge
to Smolin's hypothesis. In summary,\footnote{See Smolin (2006) for a full
explanation and critique of Vilenkin's argument.} Vilenkin's argument is as
follows: In the far future of an eternal de Sitter space-time, black holes will
be spontaneously created by the fluctuations of quantum fields, at a constant
rate. A universe such as ours appears to be, with a non-zero positive
cosmological constant $\Lambda$, will evolve towards just such a de Sitter
space-time, and the black hole production from this mechanism will dominate
that produced by astrophysical processes. Moreover, the black hole production
rate from this mechanism is proportional to the value of the cosmological
constant $\Lambda$, hence in a population of universes evolving to maximize
black hole production, the value of $\Lambda$ will be maximized. The small
value of $\Lambda$ in our own universe, argues Vilenkin, is therefore
inconsistent with cosmological natural selection.

The success of Vilenkin's argument depends largely upon the reality, or
otherwise, of the proposed black hole creation mechanism in the far future of a
de Sitter space-time, and its purported dependence upon $\Lambda$. Vilenkin's
argument currently rests upon empirically unverified, and theoretically
controversial physics, but nevertheless constitutes a serious potential problem
for cosmological natural selection.

For the purpose of this paper, however, the primary problem with Smolin's
scenario is that it cannot explain why our universe is relativistic rather than
non-relativistic, and it cannot explain why our universe is a quantum universe
rather than a classical universe, because the occurrence of black holes
requires a relativistic universe, and the occurrence of a `bounce' inside the
horizon of a black hole requires a quantum universe.

Thus, Smolin's hypothesis depends upon the assumption that there is a quantum
relativistic universe at the outset. One can ask for an explanation of why
there should be such a universe, rather than a universe in which, say,
Newtonian gravity governs the large-scale structure of space-time, or in which
classical mechanics and classical field theories govern the behaviour of any
particles and fields which exist. The existence of a quantum relativistic
universe seems to be contingent rather than necessary. There is, therefore, a
need to explain the existence of a quantum relativistic universe.

A proposal for just such an explanation will be made below. The proposal will
be expressed in terms of a variation in the value of the dimensionless
gravitational parameter $\omega$, and a variation in the value of two of the
fundamental dimensional `constants', the speed of light $c$ and Planck's
constant\footnote{Strictly, this is the `reduced' Planck constant, $\hbar =
h/2\pi$.} $\hbar$. As Kragh (2006) recounts, there is already a significant
history of such proposals in the physics community, ranging from Dirac's
hypothesis of a time-varying gravitational constant $G$, to more recent
proposals for variable speed of light (VSL) cosmologies, such as that proposed
by Albrecht and Magueijo (1999). Kragh reports out that ``hundreds of papers
have been written within the class of VSL," (p731) and claims that Magueijo's
(2003) invited review article in \emph{Reports on Progress in Physics},
indicates that the subject ``is considered exciting as well as belonging to
mainstream, if not necessarily orthodox physics," (p732).

There remains, however, considerable disagreement in the physics community over
whether a postulated variation in the value of the fundamental
\emph{dimensional} constants is well-defined or operationally meaningful, hence
the next section will be devoted to a discussion of this issue.

\section{Dimensional and dimensionless constants}

The fundamental constants of physics, $c$, $\hbar$ and $G$, are
\emph{dimensional} constants in the sense that they possess physical
dimensions, and their values must be expressed relative to a choice of physical
units. Recall in this context that there are three fundamental physical
dimensions: length [L], time [T], and mass [M]. Each physical quantity is
represented to have dimensions given by some combination of powers of these
fundamental dimensions, and each value of a physical quantity is expressed as a
multiple of some chosen unit of those dimensions. The speed of light has
dimensions of $[L][T]^{-1}$, and in CGS (Centimetre-Gramme-Second) units has
the value $c \approx 3 \times 10^{10} \; cm \, s^{-1}$; Planck's constant has
the value $\hbar \approx 10^{-27}g \; cm^2s^{-1}$ in CGS units; and Newton's
gravitational constant has the value $G \approx 6.67 \times 10^{-8} cm^3 g^{-1}
s^{-2}$ in CGS units.

The laws of physics define the necessary relationships between dimensional
quantities. The values of these quantities are variable even within a fixed
system of units, hence the lawlike equations can be said to define the
necessary relationships between dimensional \emph{variables}. Nevertheless, the
laws of physics also contain dimensional \emph{constants}. In particular, the
fundamental equations of relativity and quantum theory, such as the Einstein
field equation, the Maxwell equation, the Schr\"odinger equation, and the Dirac
equation, contain the fundamental dimensional constants, $c$, $\hbar$, and $G$.

Ultimately, dimensional constants are necessary in equations which express the
possible relationships between physical variables, because the dimensional
constants change the units on one side of the equation into the units on the
other side. As an example, consider the most famous case in physics, $E=mc^2$.
This equation can be seen as expressing a necessary relationship between the
energy-values and mass-values of a system. In CGS units the energy is in ergs,
where an erg is defined to equal one $g \; cm^2 s^{-2}$, and the mass is in
grammes. To convert the units of the quantity on the right-hand-side of the
equation into the same units as the quantity on the left-hand-side, the mass is
multiplied by the square of the speed of light in vacuum, which has units of
$cm^2 s^{-2}$. One might argue that the reason why the (square root of) the
conversion factor should be $\approx 3 \times 10^{10}$ in CGS units, rather
than any other number, follows from the definition of the $cm$ and the $s$.
Like all dimensional quantities, the value of fundamental constants such as $c$
changes under a change of physical units.

Intriguingly, the fundamental dimensional constants can also, heuristically at
least, be used to express the limiting relationships between fundamental
theories. Thus, classical physics is often said to be the limit of quantum
physics in which Planck's constant $\hbar \rightarrow 0$, and non-relativistic
physics is often said to be the limit of relativistic physics in which the
speed of light in vacuum $c \rightarrow \infty$. The flip side of this coin is
that $\hbar$ is said to set the scale at which quantum effects become relevant,
and $c$ is said to set the speeds at which relativistic effects become
relevant.

A system with action $A$ is a quantum system if the dimensionless ratio
$A/\hbar$ is small. If this ratio is large, then the system is classical. As
$\hbar \rightarrow 0$, $A/\hbar$ becomes large even for very small systems,
hence classical physics is said to be the limit of quantum physics in which
$\hbar \rightarrow 0$. Similarly, $c$ sets the speeds at which relativistic
effects become relevant in the sense that a system with speed $\nu$ is
relativistic if the dimensionless ratio $\nu/c$ is close to $1$. If the ratio
is a small fraction, then the system is non-relativistic. As $c \rightarrow
\infty$, $\nu/c$ becomes a small fraction even for very fast systems, hence
non-relativistic physics is said to be the limit of relativistic physics in
which $c \rightarrow \infty$. In a similar manner, $G$ sets the scale of
gravitational forces, and determines whether a system is gravitational or not.

Duff, however, argues that no objective meaning can be attached to a variation
in the values of the dimensional constants. According to Duff, ``the number and
values of dimensional constants, such as $\hbar, c, G, e, k$ etc, are quite
arbitrary human conventions. Their job is merely to convert from one system of
units to another\ldots the statement that $c = 3 \times 10^8 \; m/s$, has no
more content than saying how we convert from one human construct (the meter) to
another (the second)," (2002, p2-3).

To understand Duff's point, consider `geometrized' units, in which the speed of
light is used to convert units of time into units of length. Thus $c \cdot s$,
for example, is a unit of length defined to equal the distance light travels in
one second. If time is measured in units of length, then all velocities are
converted from quantities with the dimensions $[L][T]^{-1}$ to dimensionless
quantities, and in particular, the speed of light acquires the dimensionless
value $c = 1$. In geometrized units, anything which has a speed $\nu$ less than
the speed of light has a speed in the range $0 \leq \nu_{geo} < 1$:
$$ \nu_{geo} =\frac{\nu_{cgs}}{c_{cgs}} \;.
$$ Thus, the speed of light can be used to convert between velocities expressed in CGS and geometric
units as follows:

$$
\nu_{cgs} = \nu_{geo} \cdot c_{cgs} \;.
$$ Similarly, in geometrized units, the gravitational constant $G$ converts units
of mass to units of length. In fact, in geometrized units all quantities have
some power of length as their dimensions. In general, a quantity with
dimensions $L^nT^mM^p$ in normal units acquires dimensions $L^{n+m+p}$ in
geometrized units, after conversion via the factor $c^m(G/c^2)^p$, (Wald 1984,
p470).\footnote{Some interpretations of relativity hold that the unification of
space and time into space-time, entails that length $[L]$ and time $[T]$ are
simply the same dimension, $[L]=[T]$. Under these interpretations, $c=3 \times
10^{10} \; cm/s$ is seen as a conversion factor between units of the same
dimension. However, as Flores (2007) points out, ``one can consistently use
units in which $c = 1$ and hold that there is nevertheless a fundamental
distinction between space and time as dimensions."}

Whilst in geometrized units, $c=G=1$, if one changes to so-called `natural
units' (such as Planck units), then $\hbar = c = G = 1$, and these constants
disappear from the fundamental equations. Theories expressed in these natural
units provide a non-dimensional formulation of the theory, and the dimensional
variables therein becomes dimensionless. Duff, for example, points out that
``any theory may be cast into a form in which no dimensional quantities ever
appear either in the equations themselves or in their solutions," (2002, p5).
Whilst in geometrized units, all quantities have dimensions of some power of
length $[L]^n$, in Planck units all quantities are dimensionless, as a result
of division by $l_P^n$, the $n$-th power of the Planck length $l_P = \sqrt{G
\hbar/c^3} \approx 1.616 \times 10^{-33} \; cm\;.$ In particular, in natural
units all lengths are dimensionless multiples of the Planck length.

The existence of theoretical formulations in which the dimensional constants
disappear, is duly held to be one of the reasons why a postulated variation in
the values of the dimensional constants cannot be well-defined. However, whilst
different choices of units certainly result in different formulations of a
theory, and whilst the dimensional constants can indeed be eliminated by a
judicious choice of units, it should be noted that the most general formulation
of a theory and its equations is the one which contains the symbols denoting
the dimensional constants as well as the symbols denoting the dimensional
variables.

Whilst the arguments recounted above are to the effect that variations in the
fundamental dimensional constants cannot be well-defined, these arguments are
often conflated or conjoined with arguments that such changes are not
\emph{operationally} meaningful. In the latter case it is argued that a change
in a dimensional constant cannot be unambiguously measured because there is no
way of discriminating it from a change in the units of which that constant is a
multiple. For example, if the length of a physical bar, stored at a
metrological standards institute, is used to define the unit of length, one
might try to measure a change in the speed of light from a change in the time
taken for light to travel such a length. In such a scenario, it could be argued
that it is the length of the bar which has changed, not the speed of light.

Whilst it is indeed true that a change in the value of a dimensional variable
could be explained by a change in one's standard units, this is a truth which
applies to the measurement of dimensional \emph{variables}, just as much as it
applies to the measurement of dimensional \emph{constants}. The logical
conclusion of this line of argument is that only dimensionless ratios of
dimensional quantities can be determined by measurement; individual lengths,
times and masses would not be determinable, only ratios of lengths, ratios of
times, and ratios of masses. Duff duly follows this line of reasoning to its
logical conclusion, asserting that ``experiments measure only dimensionless
quantities," (2002, p5).

However, unless the dimensional quantity being measured is itself used to
define the units in which the quantity is expressed, the question of whether
one can discriminate a change in a dimensional quantity from a change in the
units of which that quantity is a multiple, is an empirical-epistemological
question rather than an ontological question. Whilst the value of a dimensional
constant does indeed change under a change of units, so does the value of a
dimensional variable, and there is no reason to infer from this that a
dimensional variable is merely a human construct. For example, the rest-mass
energy of a system changes under a change from $MeV$ to $keV$, but this is no
reason to conclude that rest-mass energy is a human construct. Hence, the
question of operational meaning may be something of a red-herring.

It is, however, certainly true that the units of time and length can themselves
be defined as functions of the fundamental dimensional constants. Thus, the
standard unit of time is defined in terms of the frequency $\nu$ of hyperfine
transitions between ground state energy levels of caesium-133 atoms:

$$
\nu = \frac{m_e^2 c^{-2}e^8}{h^5 m_N} \equiv T^{-1}\;,
$$ where $e$ is the charge of the electron, $m_N$ is the mass of the neutron,
and $m_e$ is the mass of the electron. The period of any cyclic phenomenon is
the reciprocal of the frequency, $1/\nu$, and in 1967 the second was defined in
the International System (SI) of units to consist of 9,192,631,770 such
periods.

From 1960 until 1983, the SI metre was defined to be 1,650,763.73 wavelengths
(in vacuum) of the orange-red emission line of krypton-86. This is determined
by the Rydberg length $R_\infty$:

$$
4 \pi R_\infty = \frac{m_e e^4}{ch^3} \equiv L \;.
$$ As Barrow and Tipler (1986) comment, ``if we adopt L and T as our standards of length and time then they are \emph{defined} as
constant. We could not measure any change in fundamental constants which are
functions of L and T," (p242). Since 1983 the metre has been defined in terms
of the unit of time, the second, so that a metre is defined to be the distance
travelled by light, in a vacuum, during $1/299 792 458$ of a second. Such
considerations lead Ellis (2003) to claim that ``it is\ldots not possible for
the speed of light to vary, because it is the very basis of measuring
distance."

Magueijo and Moffat (2007) acknowledge that if the unit of length is defined in
such a manner, then the constancy of the speed of light is indeed a tautology.
However, they then provide the following riposte: ``An historical analogy may
be of use here. Consider the acceleration of gravity, little $g$. This was
thought to be a constant in Galileo's time. One can almost hear the Ellis of
the day stating that $g$ cannot vary, because `it has units and can always be
defined to be constant'. The analogy to the present day relativity postulate
that $c$ is an absolute constant is applicable, for the most common method for
measuring time in use in those days did place the constancy of $g$ on the same
footing as $c$ nowadays. If one insists on defining the unit of time from the
tick of a given pendulum clock, then the acceleration of gravity is indeed a
constant by definition. Just like the modern speed of light $c$. And yet the
Newtonian picture is that the acceleration of gravity varies," (p1-2).

Whilst there is considerable disagreement that the values of fundamental
dimensional constants have any theoretical significance, there is a consensus
that each different value of a fundamental dimensionless constant, such as the
fine structure constant $\alpha = e^2/\hbar c$, defines a different theory. The
values of the dimensionless constants are, by definition, invariant under any
change of units, they remain obstinately in the dimensionless formulation of a
theory, and their values have to be set by observation and measurement.
Dimensionless constants, however, are themselves merely functions $f(c, \hbar,
G)$ of dimensional constants, in which the dimensions of the units cancel. If
the variation of dimensionless constants is meaningful, and if dimensionless
constants are functions of the dimensional constants, then one might ask how
variation in the former can be achieved without variation in the latter. As
Magueijo (2003) comments: ``If $\alpha$ is seen to vary one cannot say that all
the dimensional parameters that make it up are constant. Something - $e$,
$\hbar$, $c$, or a combination thereof - has to be varying. The choice amounts
to fixing a system of units, but that choice has to be made\ldots In the
context of varying dimensionless constants, that choice translates into a
statement on which dimensional constants are varying."

Kragh claims that ``Magueijo and Albrecht were aware of [objections such as
those which Duff later raised] in their 1999 paper, where they argued that
physics necessarily involves dimensional quantities and that a time variation
of these can be determined on grounds of conventionalism. Moreover, they
pointed out that although it is a matter of convenience to decide which
dimensional quantities are variable and which are constant, the choice has
physical implications as it will typically lead to different predictions,"
(2006, p734). (However, if such choices do indeed lead to different
predictions, then such a choice would involve more than a matter of mere
convention).

We will see in the next section how the proposed variation in the dimensional
constant $G$ leads to a generalisation of general relativity containing a
dimensionless parameter $\omega$, whose limit $\omega \rightarrow \infty$
corresponds to general relativity. There is, as yet, no comparable
generalisation of relativistic quantum theory, hence the questions of stability
under deformation which we consider below, must first be evaluated for the
dimensional constants $c$ and $\hbar$. Thus, in the presentation below for
extending cosmological natural selection, the proposal will be expounded in
terms of the dimensional constants $c$ and $\hbar$, and the dimensionless
parameter $\omega$.

\section{Stable mathematical structures}

As a first step to explaining the type of universe population postulated by
Smolin, the following conjecture is proposed:

\begin{Conjecture}At some level, the structure of our physical universe is a stable mathematical structure.
\end{Conjecture}

A stable (`rigid') mathematical structure is a structure for which any
deformation, in some specified class of deformations, merely leads to an
isomorphic structure (see Mazur 2004). A deformation is a continuous variation
of a structure by means of some parameter(s). Intriguingly, some of the most
fundamental structures which describe our universe are, indeed, stable
structures (Faddeev 1991; Vilela Mendes 1994).

Firstly, whilst the Lie algebra of the inhomogeneous Galilei group, the local
symmetry group of Galilean relativity, is an unstable structure, it deforms
into a family of Lie algebras, parameterised by the speed of light $c$. All of
these 10-dimensional Lie groups are mutually isomorphic to the Poincare group,
the local space-time symmetry group of general relativity. This family of Lie
groups transforms into the Galilei group in the limit $c \rightarrow \infty$.

Secondly, whilst the Lie algebra defined by the Poisson bracket on the space of
observables in a classical physical theory is an unstable Lie algebra, it
deforms into a family of Lie algebras, parameterised by Planck's constant
$\hbar$. All of these Lie algebras are mutually isomorphic to the Lie algebra
defined by the commutator on the space of observables in the corresponding
quantum theory. If one thinks of each value of $\hbar$ as defining a different
quantum theory, then this amounts to the deformation of a classical theory into
a family of quantum theories. The same type of deformation can be performed
using C$^*$-algebras: ``the classical algebra of observables is `glued' to the
family of quantum algebras of observables in such a way that the classical
theory literally forms the boundary of the space containing the pertinent
quantum theories (one for each value of $\hbar > 0$)," (Landsman 2005, Section
4.3). The family of quantum theories transforms into classical theory in the
limit $\hbar \rightarrow 0$.

At least some of the parameters of physics are therefore the deformation
parameters of mathematical structures, and a relativistic quantum universe,
such as our own, corresponds in at least some respects to a stable structure.

There are also some suggestive facts from the standard model of
particle physics, where each gauge force field has an `internal'
symmetry group, called the gauge group. A gauge group must be a
compact, connected Lie Group. In our universe, the gauge group of
the electromagnetic force is $U(1)$, the gauge group of the
electroweak force is $U(2) \cong SU(2) \times U(1)/\mathbb{Z}_2$,
and the gauge group of the strong force is $SU(3)$. Now, the
vanishing of the second cohomology group of a Lie algebra entails
that the Lie algebra is stable (see Vilela Mendes 1994).
Semi-simple Lie algebras have a trivial second cohomology group,
hence semi-simple Lie algebras are stable structures. Every simple
Lie algebra is semi-simple, and $SU(2)$ and $SU(3)$ are simple Lie
groups, hence the Lie algebras of $SU(2)$ and $SU(3)$ are stable
structures. Moreover, the Lie algebra of $U(1)$ is $\mathbb{R}$,
and, as the only 1-dimensional real Lie algebra, this is also a
stable Lie algebra.

There are, however, many simple, compact, connected Lie groups. The list of the
simply connected ones alone, contains the special unitary groups $SU(n), \; n
\geq 2$; the symplectic groups\index{symplectic groups} $Sp(n), \; n \geq 2$;
the spin groups\index{spin group} $Spin(2n+1), \; n \geq 3$; the spin groups
$Spin(2n), \; n \geq 4$; and the five exceptional Lie groups\index{exceptional
groups} $E_6$, $E_7$, $E_8$, $F_4$, and $G_2$, (Simon 1996, p151). Hence,
structural stability alone can only go so far towards explaining why the gauge
fields which exist in our universe are those which have either $U(1)$, $U(2)
\cong SU(2) \times U(1)/\mathbb{Z}_2$, or $SU(3)$ as their gauge groups. The
gauge fields which exist in our universe might have to be explained by a
combination of structural stability and additional constraints on the
permissible gauge fields.

To reiterate, a stable structure is defined to be a structure which remains
isomorphic `under a specified class of deformations'. Hence, whether or not a
structure is stable depends upon the class of deformations under consideration,
and the proposal that our physical universe is a stable mathematical structure
is only meaningful with respect to a designated class of deformations. This
requirement is supplied by the next proposal, which postulates that there is a
physical process which randomly changes the parameters of physics:

\begin{Conjecture}There is a physical process which randomly changes deformation parameters such as $c$ and $\hbar$.
\end{Conjecture}

The proposal, then, is that our physical universe is a stable structure with
respect to the class of deformations corresponding to this physical process.

Despite the difficulties of defining or unambiguously ascertaining by
observation and measurement whether the dimensional parameters of physics are
actually subject to variation, the proposal above constitutes a potentially
testable conjecture, and is therefore a scientific conjecture. The existence of
such a physical process will inevitably result in a relativistic quantum
universe, even if it started with a classical universe, or a non-relativistic
universe. Moreover, with the imposition perhaps of further constraints, such a
process might produce a universe with gauge fields like our own, even if it
started with quite different gauge fields. If so, then a quantum relativistic
universe with the gauge force fields we observe, would be a stable region in
the mathematical `landscape'.

However, such a conjecture only goes so far; the mathematical structures which
describe our universe can only be cast as stable structures at a quite general
level. Whilst a quantum relativistic universe can be said to be a stable
structure, the specific structures of the particles and fields in such a
universe cannot. For example, the coupling constant of a gauge field with gauge
group $G$ corresponds to a choice of metric in the Lie algebra $\mathfrak{g}$,
(Derdzinksi 1992, p114-115), and the particular metrics chosen in our own
universe are not stable in any sense; different coupling constants correspond
to non-isometric structures in the gauge group Lie algebras.

Thus, to explain the detailed mathematical structure of our universe, the
notion of evolution towards stable mathematical structures must be combined
with Smolin's scenario of cosmological evolution by natural selection:

\begin{Conjecture}Our universe belongs to a population of quantum relativistic universes, evolving by natural
selection.
\end{Conjecture}

To reiterate, Smolin's scenario explains how parameters of the standard model,
such as the coupling constants of the gauge fields, come to possess the values
we observe. It is proposed in this paper that the evolution of universes which
occurs within Smolin's scenario, takes place within a context established by
the prior evolution of a stable mathematical structure, at a more general level
than the level at which the natural selection process operates. In fact, the
evolution of universes in Smolin's scenario is \emph{dependent} upon the prior
evolution of a quantum relativistic structure. It is proposed in this paper
that there are random processes which deformed the structure of the universe,
or a region thereof, into a quantum relativistic universe, and thereon, the
processes postulated in Smolin's evolution by natural selection produced a
multiverse of quantum relativistic universes.

The other fundamental parameter of physics, along with $c$ and $\hbar$, is the
gravitational constant $G$. To introduce this into the theory of cosmological
evolution, the Brans-Dicke theory of gravitation can be deployed. Brans-Dicke
theory is a generalisation of general relativity, in which the reciprocal of
the gravitational constant $1/G$, is replaced by a scalar field $\phi$. The
scalar field $\phi$ is an effective (reciprocal of the) gravitational constant,
which is capable of varying from place to place, and from time to time. The
Einstein field equations of general relativity generalise to the Brans-Dicke
field equations, in which the combination of the matter stress-energy tensor
$T$ and the scalar field $\phi$ generate the metric tensor. These field
equations contain a dimensionless parameter $\omega$ called the Brans-Dicke
coupling constant. For each different value of $\omega$, there is a different
Brans-Dicke theory.

The value of $\omega$ has to be set by experiment and observation, and current
astronomical observations have established a lower bound such that $\omega >
40,000$ (Bertotti \emph{et al} 2003). General relativity is obtained from the
family of Brans-Dicke theories in the limit $\omega \rightarrow \infty$ (with
some exceptions; see Faroni 1999). This means that general relativity is
unstable in the space of mathematical structures. A deformation of general
relativity takes it into the space of Brans-Dicke theories. Hence, if we
postulate that $\omega$ is subject to the same random variation to which $c$
and $\hbar$ are subject, then from the theory of cosmogenic drift we obtain the
prediction:

\begin{Conjecture}The correct theory of classical gravitation in our universe is the Brans-Dicke theory,
for some value of $\omega$.
\end{Conjecture}

Whilst a quantum relativistic universe is a stable universe, it seems that a
strictly general relativistic universe is unstable. We are currently unable to
place a finite upper bound on the value of $\omega$ in our universe, hence we
are currently unable to observationally distinguish our Brans-Dicke universe
from a general relativistic universe. This, however, may simply be a result of
the inadequacy of current observational technology; after all, the current
lower bound on $\omega$ is $40,000$, which is still a long way from $\infty$.

A population of universes in which gravity is governed by Brans-Dicke theory,
is still a population of relativistic universes, and, crucially, black holes
exist within all the Brans-Dicke theories. Exact vacuum solutions of the
Einstein field equations, supplemented by the addition of a scalar field which
is such that $\phi = 1$ everywhere, become exact vacuum solutions of any
Brans-Dicke theory. Hence, a population of Brans-Dicke universes can evolve by
natural selection just as much as a population of general relativistic
universes.

\section{Cosmogenic drift}

In evolutionary biology it is known that evolution by natural selection is not
the only important evolution process, and that in the absence of selection
pressures, the evolution of a population will be dominated by random variations
in the genome, a process called genetic drift. Similarly, the proposal made
here suggests that the values of the parameters of physics cannot be wholly
explained by cosmological evolution by natural selection. However, whilst
genetic drift is a process which applies to a population of biological entities
reproducing with inheritance and random mutation, the cosmological process
postulated here is not restricted to reproducing entities, and in particular is
postulated as a necessary prelude to the creation of a population of
reproducing universes. Moreover, genetic drift does not couple the idea of
random variation to the notion of stable structures. Nevertheless, one might
wish to refer to the postulated process which produced a quantum relativistic
universe as \emph{cosmogenic drift}.

To best explain cosmogenic drift, we shall need a concept from evolutionary
biology known as the \emph{fitness landscape}. Each point on this landscape
corresponds to a different combination of genes, and the height of the
landscape at each point represents the average number of progeny produced by an
organism with that combination of genes, which themselves survive to reproduce.
The height of the landscape therefore represents the `fitness' of each possible
genotype. Each progenitor produces offspring with genomes in a small
neighbourhood of the position of the progenitor in the landscape. In those
parts of the landscape where selection pressures are weak, none of the progeny
will have a greater fitness. When selection pressures are weak, the fitness
landscape is therefore almost flat. Evolution of a biological population across
a flat part of the fitness landscape will be driven by random diffusion. In
contrast, in those parts of the landscape where selection operates, the
landscape will possess gradient. In these parts of the landscape, some of the
progeny produced within a small neighbourhood of one genotype will lie at a
slightly greater height because they yield a greater number of progeny which
themselves survive to reproduce. As a consequence, the population will come to
be dominated by this new genotype, and will take a step-up to a slightly
greater height in the fitness landscape. This is biological evolution by
natural selection.

Smolin (1997) suggested that there is a cosmic fitness landscape analogous to
the biological one, with each point corresponding to a combination of values
for the parameters of physics, and the height at each point representing the
number of progeny produced by a universe with that combination of parameters.
This proposal can be extended by postulating that the cosmic fitness landscape
has a lowest level plateau, corresponding to all the possible types of universe
which do not reproduce. Each point in the lowest level of the cosmic landscape
has zero height because none of these universes yield any progeny, and the
landscape is flat here because natural selection cannot operate in the absence
of reproduction. Evolution does, nevertheless, occur in this part of the cosmic
landscape. Universes, it is proposed, evolve by random diffusion in flat parts
of the cosmic fitness landscape; in particular, universes which cannot
reproduce, inhabiting the lowest level of the landscape, evolve by such
cosmogenic drift. Eventually, a universe undergoing random cosmogenic drift
will evolve into a quantum relativistic universe, a universe type which is
capable of reproduction. The part of the cosmic fitness landscape which
contains universes capable of reproduction, corresponds to a region of
precipitous elevation in the landscape. This part of the landscape possesses a
variety of gradients, and evolution by natural selection operates, as suggested
by Smolin, in the same manner that it operates in the biological fitness
landscape.

There is no necessity for cosmogenic drift to be the type of evolution which
falls under the aegis of the Lagrangian-Hamiltonian dynamics of conventional
physical theory. It is true that some of the modern VSL cosmologies replace the
constant speed of light $c$ with a scalar field $\psi = c(x)$, and propose a
modified Lagrangian incorporating the Lagrangian $\mathscr{L}_{\psi}$ of that
scalar field, and the Brans-Dicke theory is indeed obtained by replacing $G$
with a scalar field which results in a modified Lagrangian. Nevertheless, it is
not proposed that the dimensionless gravitational constant $\omega$ evolves
according to Lagrangian-Hamiltonian dynamics. Nor, if there is a generalisation
of quantum relativistic theory which contains comparable dimensionless
parameters, is it proposed that these parameters would evolve according to
Lagrangian-Hamiltonian dynamics. The type of evolution proposed here is pure
random diffusion, and has nothing to do with the Lagrangian-Hamiltonian
dynamics of any quantum relativistic field $\psi$, for the evolution of a
quantum relativistic universe is itself the proposed outcome of this process.
As Magueijo and Moffat (2007) point out, ``it is not true that a theory has to
be defined by a Lagrangian or a Hamiltonian\ldots Absence of a Lagrangian
formulation is far from being a general feature of VSL, but we argue that it
may be the point of those that attack the philosophical foundations of physics
at its most fundamental level, introducing the concept of intrinsic evolution
in the laws of physics," (p4).

Random diffusion is a type of \emph{stochastic process}, so if the theory of
cosmogenic drift is to be further developed, and if observable predictions are
to be derived from it, it will be necessary to employ the mathematics of
stochastic processes, a brief explanation of which is duly required.

Mathematicians define a stochastic process to be a time-ordered family of
random variables $X_t$ upon a probability space $\Omega$. Whilst this is not
particularly illuminating in itself, the implicit idea is that $\Omega$ is the
path-space for the system under consideration. In other words, each point in
this probability space, $\omega \in \Omega$, represents a possible history of
the system.\footnote{In the case of a stochastic process, these histories will
typically be non-differentiable.}

By definition, a random variable $X$ is a function on a probability space
$\Omega$ which possesses a probability distribution over its range of possible
values, by virtue of the probability measure on the subsets of the probability
space $\Omega$. In the case of a stochastic process, $X_t$ is a function on the
path-space of the system which represents the position of the system at time
$t$.\footnote{`Position' here can be taken to be spatial position, or any sort
of state-defining value, such as the price of a financial stock.} Thus
$X_t(\omega)$, the value of the random variable $X_t$ at the point $\omega \in
\Omega $, is the position of the system at time $t$ in history $\omega$. $X_t$
takes different values at different points because the different points in
$\Omega$ correspond to different histories of the system. The probability
measure on $\Omega$, the space of histories, determines the probability
distribution over the range of each random variable $X_t$, and thereby
determines a probability distribution over position at each time $t$. Different
positions at time $t$ have different probabilities because different histories
have different probabilities.

A stochastic process can also be defined by a function $G(x,x'; t)$ which
specifies the probability of a transition from $x$ to $x'$ over a time interval
$t$. Given an initial probability distribution $\rho(x,0)$, this determines the
probability distribution $\rho(x',t)$ at a future time $t$:

$$\rho(x',t) = \int G(x,x'; t)\rho(x,0) \; dx \,.$$

In fact, given the transition probabilities $G(x,x'; t)$ and an initial
probability distribution $\rho(x,0)$, a probability measure on the path-space
$\Omega$, and the time evolution of the probability distribution $\rho(x,t)$,
are both determined.

In the special case of a discrete stochastic process, with the transition
probability of going from $x$ to $y$ in one time-step denoted as $T(x,y)$, the
probability $p(\gamma)$ of a path $\gamma$ defined by the sequence of positions
$(x_0,\ldots,x_n)$ is defined to be:

$$p(\gamma) = T(x_{n-1},x_n)\ldots T(x_0,x_1)\rho(x_0,0)\;.$$

As a stochastic process, random diffusion comes in a number of different
varieties, so the first question one might pose to the cosmogenic drift
hypothesis, is to ask which specific type of diffusion is postulated to
operate. The simplest type of diffusion is Brownian motion, (also termed a
Wiener process), which is a simple random walk in which the increments between
random variables $S_t$ have a normal distribution with a mean value of zero.
Geometric Brownian motion, in contrast, is such that each random variable $S_t$
has a lognormal distribution. Moreover, Brownian motion and geometric Brownian
motion can each possess a drift, which ensures that the mean values of the
random variables $S_t$ evolve as if under the action of an external force. For
example, the Black-Scholes equation, used to calculate the price of financial
options, assumes that the value of underlying stocks will evolve according to
geometric Brownian motion with drift. This stochastic process is typically
denoted as

$$dS_t = \mu S_tdt + \sigma S_t dW_t \;,$$ where $\mu$ is the percentage drift due to the expected risk-free
rate-of-return on the underlying stock, $\mu S_t$ is the drift-rate, $\sigma$
is the volatility of the stock, $\sigma S_t$ is the diffusion rate, and $dW_t$
is a standard Wiener process.

In terms of the probability distribution $\rho$ on the range of the random
variables $S_t$, in the case of simple Brownian motion it evolves according to
the diffusion equation,

$$\partial \rho/\partial t = D \nabla^2 \rho \;,$$ whilst in the case of Brownian motion with drift tending towards a terminal
drift-rate $\nu$, it evolves according to the diffusion equation with drift:

$$\partial \rho/\partial t = D \nabla^2 \rho - \nu \nabla \rho \;.$$ $D$ here is the so-called diffusion coefficient, $\nabla^2$ is the Laplacian,
and $\nabla$ is the gradient.

To explain the prior evolution of a quantum relativistic universe, diffusion
seems to work equally well as diffusion-with-drift. Simple random diffusion
will eventually evolve a sterile universe into a quantum relativistic universe,
at which point reproduction will be triggered, and evolution by natural
selection can kick-in. Diffusion-with-drift towards the relevant part of the
cosmic fitness landscape will produce a quantum relativistic universe in a
shorter time-scale, but given a presumably eternal length of time, a sense of
urgency seems unnecessary. The form of the stochastic process distribution
$\rho$, whether it is normal, lognormal, or otherwise, is also largely
unimportant to the outcome, (although if negative values of the parameters are
to be excluded, then one might stipulate geometric Brownian motion). All of
which, unfortunately, seems to mitigate against the possibility of deriving
potentially observable predictions from the theory. It might still be possible
to observe the variations in the parameters of physics within our own universe,
and thence to infer the nature of the stochastic process, but that information
would then have to be fed back into the theory, rather than derived from it.

One can also ask which stochastic differential equation is satisfied by the
random variation of parameters in Smolin's scenario, and given an answer, one
could ask why cosmological natural selection utilises one particular type of
random process rather than another. However, because Smolin's scenario is
placed within the context of a quantum relativistic universe, it could be
suggested that this type of stochastic evolution is determined by the rules of
quantum gravity. In contrast, \emph{any} type of random cosmogenic drift, from
an arbitrary starting point, would eventually evolve a quantum relativistic
universe as a stable structure, and from that point, a population of
reproducing universes would ensue.

\end{document}